\documentclass[12pt,times]{article}
\title{Self-cleaning of hydrophobic rough surfaces by coalescence-induced wetting transition}

\author{Kaixuan Zhang$^{1,2,*}$
        Zhen Li$^{2,}$\footnote{The first two authors contributed equally to this work.}$^{~,}$\footnote{Correspondence email: \href{mailto:zhen_li@brown.edu}{zhen\_li@brown.edu}},
        Martin Maxey$^2$,
        Shuo Chen$^{1,}$\footnote{Correspondence email: \href{mailto:schen_tju@mail.tongji.edu.cn}{schen\_tju@mail.tongji.edu.cn}}
        \\and George Em Karniadakis$^{2}$\\
\small{$^1$School of Aerospace Engineering and Applied Mechanics, Tongji University, Shanghai 200092, China}\\
\small{$^2$Division of Applied Mathematics, Brown University, Providence, Rhode Island, 02912, USA} }

\date{}

\RequirePackage{lineno}
\usepackage[colorlinks=true]{hyperref}
\usepackage[sort&compress,numbers]{natbib}
\usepackage{caption}
\usepackage{subcaption}
\usepackage{multirow,array,color,mathrsfs,subcaption}

\usepackage{amsmath,fancyhdr,amsthm,amssymb,amsfonts,version}
\usepackage{bbm,version}
\usepackage[section]{placeins}
\usepackage{stmaryrd}
\usepackage{bm}

\usepackage{dsfont}

\usepackage[top=1.5cm,bottom=2.5cm,left=1.5cm,right=1.5cm]{geometry}

\usepackage[figuresright]{rotating}
\usepackage{extarrows}
\begin{document}

\maketitle
\vspace{-20pt}

\begin{abstract}
The superhydrophobic leaves of a lotus plant and other natural surfaces with self-cleaning function have been studied intensively for the development of artificial biomimetic surfaces. Surface roughness generated by hierarchical structures is a crucial property required for superhydrophobicity and self-cleaning. Here, we demonstrate a novel self-cleaning mechanism of textured surfaces attributed to a spontaneous coalescence-induced wetting transition. We focus on the wetting transition as it represents a new  mechanism, which can explain why droplets on rough surfaces are able to change from the highly adhesive Wenzel state to the low-adhesion Cassie-Baxter state and achieve self-cleaning. In particular, we perform many-body dissipative particle dynamics simulations of liquid droplets (with diameter $89~\mu m$) sitting on mechanically textured substrates. We quantitatively investigate the wetting behavior of an isolated droplet as well as coalescence of droplets for both Cassie-Baxter and Wenzel states.\ Our simulation results reveal that droplets in the Cassie-Baxter state have much lower contact angle hysteresis and smaller hydrodynamic resistance than droplets in the Wenzel state. When small neighboring droplets coalesce into bigger ones on textured hydrophobic substrates, we observe a spontaneous wetting transition from a Wenzel state to a Cassie-Baxter state, which is powered by the surface energy released upon coalescence of the droplets. For superhydrophobic surfaces, the released surface energy may be sufficient to cause a jumping motion of droplets off the surface, in which case adding one more droplet to coalescence may increase the jumping velocity by one order of magnitude. When multiple droplets are involved, we find that the spatial distribution of liquid components in the coalesced droplet can be controlled by properly designing the overall arrangement of droplets and the distance between them. These findings offer new insights for designing effective biomimetic self-cleaning surfaces by enhancing spontaneous Wenzel-to-Cassie wetting transitions, and additionally, for developing new non-contact methods to manipulate liquids inside small droplets via multiple-droplet coalescence.
\end{abstract}

\section{Introduction}\label{sec:int}
Wettability is a characteristic property of a solid surface, and depends on the intrinsic surface energy as well as the surface morphology. According to the value of wetting contact angle~$\theta$, the surface wetting behavior can generally be divided into four different regimes~\cite{2015Mohamed}: superhydrophilicity for $\theta < 10^\circ$, hydrophilicity for $10^\circ \le \theta < 90^\circ$, hydrophobicity for $90^\circ < \theta \le 150^\circ$ and superhydrophobicity for $\theta > 150^\circ$. Superhydrophobic surfaces exhibit very low adhesion and small sliding angles, and they are associated with anti-icing~\cite{2009Cao}, anti-fogging~\cite{2014Sun} and self-cleaning~\cite{2012Liu} properties, which have attracted increasing attention for both purely academic interest and industrial applications~\cite{2011Kubiak}.
However, the superhydrophobicity cannot be obtained for flat surfaces by just lowering the intrinsic surface energy. It was reported that a -CF$_3$ terminated surface can have the lowest free energy and the best hydrophobicity, but with a contact angle of merely $120^\circ$ on flat surfaces~\cite{1999Nishino}.
Learning from natural surfaces, i.e., lotus leaves~\cite{2002Feng}, water strider legs~\cite{2004Gao} and butterfly wings~\cite{2007Zheng}, we see that nature does not require an extremely low surface energy, such as -CF$_3$ groups, to achieve superhydrophobicity. Instead the surface morphology can play a crucial role affecting wettability of solid surfaces and enhancing hydrophobicity~\cite{2012Zhang}. Inspired by the microstructures on natural surfaces, significant efforts~\cite{2006BhushanB,2006BhushanJ,2017Liu,2009BhushanJY,2009KOCH,2009BhushanB,2017Wang_Guo} have been made to design and fabricate artificial superhydrophobic surfaces by creating biomimetic structures on hydrophobic substrates.

\begin{figure}[b!]
\centering
\includegraphics[width=0.48\textwidth]{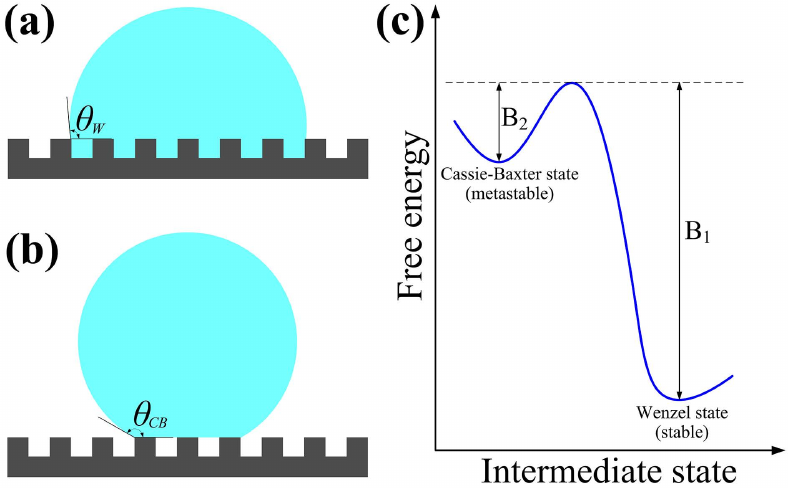}
\caption{Sketch of a liquid droplet in (a) the Wenzel state with an apparent contact angle $\theta_{\rm W}$, and (b) the Cassie-Baxter state with an apparent contact angle $\theta_{\rm CB}$; (c) schematic plot of the free energy of the two wetting states, where B$_1$ represents the Wenzel-to-Cassie transition barrier while B$_2$ the Cassie-to-Wenzel transition barrier.}
\label{fig:FreeEnergy}
\end{figure}

Two distinct hypotheses have been proposed to explain why the surface roughness of a hydrophobic solid is able to significantly enhance the hydrophobicity, with wetting contact angles increased from less than $120^\circ$ to a value as high as $175^\circ$~\cite{2003Lafuma}.
The Wenzel model~\cite{1936Wenzel} considers a homogeneous wetting state where the liquid fills in the grooves of a rough surface so that the surface area of liquid-solid interface is geometrically increased, as shown in Fig.~\ref{fig:FreeEnergy}(a).
The Cassie-Baxter model~\cite{1944Cassie} considers a heterogeneous wetting state where the droplet sits partially on the air trapped in the asperity valleys, as shown in Fig.~\ref{fig:FreeEnergy}(b).
In general, liquid droplets in the Wenzel state adhere more strongly to the rough surface with larger contact-angle hysteresis than the ones in the Cassie-Baxter state~\cite{2009Koishi}.
Therefore, in order to achieve high water repellency and the self-cleaning property, both natural and artificial superhydrophobic surfaces prefer the Cassie-Baxter state over the Wenzel state.
Given a hydrophobic solid, Fig.~\ref{fig:FreeEnergy}(c) shows a schematic plot of the free energy of the two wetting states, where $B_1$ represents the Wenzel-to-Cassie transition barrier while $B_2$ the Cassie-to-Wenzel transition barrier~\cite{2017Gong}. The free energy of the Wenzel state is lower than that of the Cassie-Baxter state with $B_1 > B_2$, meaning that the Cassie-Baxter state is metastable and a Cassie-to-Wenzel transition is more favorable than a Wenzel-to-Cassie transition.
Then, it would be a paradox because the droplets on self-cleaning rough surfaces should be in the non-sticking Cassie-Baxter state, but the sticky Wenzel state is a more favorable and stable state for hydrophobic solids.
There should exist a mechanism to change a droplet stuck in the Wenzel state to the Cassie-Baxter state.

It is known from Fig.~\ref{fig:FreeEnergy}(c) that extra energy is required to overcome the free-energy barrier $B_1$ for a Wenzel-to-Cassie wetting transition. This can be powered by different external energy sources, i.e., electrostatic field~\cite{2011Manukyan}, mechanical vibration~\cite{2009BoreykoChen}, acoustic energy~\cite{2013DR}, and impulse heating~\cite{2011LG}. Many previous works~\cite{2004Patankar,2014Murakami,2015Borma,2018Fang} have studied the wetting transition from a higher energy Cassie droplet to a lower energy Wenzel droplet. It was believed that the Cassie-to-Wenzel wetting transition is irreversible without external energy inputs, and hence a spontaneous Wenzel-to-Cassie wetting transition was considered impossible~\cite{2004Patankar,2018Fang}. Recently, the phenomenon of coalescence-induced droplet jumping on superhydrophobic surfaces has received extensive attention because the surface energy released upon coalescence can be sufficient to induce a spontaneous jumping motion of coalesced droplets. The jumping motion of coalesced droplets has been intensively investigated from different perspectives, including theoretical analyses~\cite{2011Wang,2013Lv}, experimental measurements~\cite{2013Wisdom,2014Enright,2016Chen,2017Mulroe}, and numerical simulations~\cite{2013Peng,2015Samaneh,2017Attarzadeh,2015Liang,2018Xie}.
In particular,
Wang et al.~\cite{2011Wang} proposed a theoretical model based on the energy balance of surface energy, kinetic energy and the viscous dissipation to  analyze the jumping velocity of two equal-sized droplets. This theoretical model was then improved by Lv et al.~\cite{2013Lv} by including the energy loss in deforming the droplet, which was used to explain why the critical diameter for coalesced-induced jumping in many experimental observations was of order $10~\mu m$.
Wisdom et al.~\cite{2013Wisdom} used the cicada wing as a model surface and experimentally demonstrated that a self-propelled jumping condensate on superhydrophobic surfaces can remove a variety of contaminants, including both hydrophobic/hydrophilic and organic/inorganic particles with a diameter as large as $100~\mu m$.
In the experiments performed by Mulroe et al.~\cite{2017Mulroe}, they characterized the dynamic behavior of condensing droplets jumping off a superhydrophobic surface with different surface nanostructures by varying the topography of the nanopillars, and found that the critical jumping diameter of droplets can be reduced by an order of magnitude (down to $2~\mu m$) by designing surface nanostructures.
Peng et al.~\cite{2013Peng} simulated the dynamics of droplet jumping induced by droplet coalescence on a superhydrophobic surface using the multiphase Lattice Boltzmann method (LBM), and obtained an accurate prediction of the jumping velocity of the coalesced droplet as compared with experimental measurements.
Attarzadeh and Dolatabadi~\cite{2017Attarzadeh} employed the volume-of-fluid method coupled with a dynamic contact angle model and discretized by the finite volume method to simulate the self-propelled jumping of droplets on homogeneous and heterogeneous superhydrophobic surfaces. They reported that a direct inclusion of surface topography is crucial for the correct modeling of droplet-surface interactions if the droplet size is smaller than a critical relative roughness.
Also, molecular dynamics (MD) simulations~\cite{2015Liang,2018Xie} were used to investigate the coalescence-induced jumping process of equal-sized and unequal-sized nanodroplets.

In the present paper, we construct a mesoscopic particle-based model for liquid droplets sitting on textured hydrophobic surfaces, and quantitatively investigate the wetting behavior of an isolated droplet as well as coalescence of droplets for both Cassie-Baxter and Wenzel states. For this we use many-body dissipative particle dynamics (mDPD) simulations to study the wettability of micro-pillar textured surfaces and the dynamics of coalescence-induced droplet jumping on superhydrophobic surfaces. We focus on the wetting transition and explore whether we can observe a spontaneous Wenzel-to-Cassie wetting transition powered by the surface energy released upon droplet coalescence. Our aim is to explain why droplets on rough surfaces are able to spontaneously change from the sticky Wenzel state to the non-sticky Cassie-Baxter state, without external energy inputs, to achieve self-cleaning.

The remainder of this paper is organized as follows. We first describe the details of mDPD formulation and its parameterization for wetting problems, including a clarification of setting the empirical relaxation parameter~$\lambda$ in the modified velocity-Verlet algorithm for mDPD simulations. Subsequently, we present the numerical results of the wetting behavior of droplets on rough surfaces for both Cassie-Baxter and Wenzel states, and coalescence-induced wetting transition of droplets on mechanically textured substrates. Moreover, we discuss the coalescence dynamics of multiple droplets. Finally, we conclude with a brief summary and discussion.

\section{Numerical Method}
\subsection{Many-body DPD model}
A dissipative particle dynamics (DPD) system consists of many interactive particles, that are coarse-grained and each represents the collective behavior of a group of actual fluid molecules. The time evolution of a typical DPD particle~$i$ with unit mass~$m_i \equiv 1$ is governed by Newton's equation of motion, which is described as~\cite{1997Groot}
\begin{align}
\frac{\mathrm{d}^2 \mathbf{r}_i}{\mathrm{d} t^2} = \frac{\mathrm{d} \mathbf{v}_i}{\mathrm{d} t}=\mathbf{F}_{i}=\sum_{i\neq j}(\mathbf{F}_{ij}^{C}+\mathbf{F}_{ij}^{D}+\mathbf{F}_{ij}^{R}) \ ,
\label{equ:gov1}
\end{align}
where~$t$ denotes time, $\mathbf{r}_i$,~$\mathbf{v}_i$ and~$\mathbf{F}_i$ represent position, velocity and force vectors, respectively. The total force $\mathbf{F}_i$ takes into account three pairwise components, i.e., the conservative force~$\mathbf{F}_{ij}^{C}$, the dissipative force~$\mathbf{F}_{ij}^{D}$ and the random force~$\mathbf{F}_{ij}^{R}$.

In the classical DPD model, the conservative force is purely repulsive and is computed by~$\mathbf{F}^C_{ij}=a w_C(r_{ij})\mathbf{e}_{ij}$, where~$a$ is a positive parameter, $r_{ij}=|\mathbf{r}_{ij}|$ with~$\mathbf{r}_{ij}=\mathbf{r}_i-\mathbf{r}_j$ is the distance between particles $i$ and $j$, and~$\mathbf{e}_{ij}=\mathbf{r}_{ij}/r_{ij}$ is the unit vector. The weight function~$w_C(r)$ usually takes the simple form~$w_C(r)=1-r/r_c$, where~$r_c$ is a cutoff radius beyond which~$w_C(r)$ vanishes. Groot and Warren~\cite{1997Groot} computed the equation of state (EOS) of a DPD fluid, which is a quadratic function~$p \approx \rho k_BT + \alpha a \rho^2$ with a coefficient~$\alpha = 0.101\pm0.001$. This monotonic function as an EOS does not contain a van der Waals loop and cannot model phenomena involving vapor-liquid coexistence or free-surfaces of single component fluids.
To this end, Pagonabarraga and Frenkel~\cite{2001Pagonabarraga} introduced many-body interactions in DPD, so that the conservative force between a pair of DPD particles depends not only on their relative position but also on their local densities. This defines an extension of DPD, named the many-body DPD (mDPD) model.
Subsequently, Warren~\cite{2003Warren} proposed a density-dependent conservative force given by
\begin{equation}
\mathbf{F}^{C}_{ij} = A w_{c}(r_{ij}) \mathbf{e}_{ij} + B (\rho_{i} + \rho_{j})w_d(r_{ij})\mathbf{e}_{ij},
\label{eq:FC}
\end{equation}
where the first term with~$A < 0$ represents an attractive force with a cutoff radius~$r_c$, and the second term with~$B > 0$ is a density-dependent repulsive force with a cutoff radius~$r_d$. The weight functions are defined as~$w_c(r)=1-r/r_c$ and~$w_d(r)=1-r/r_d$. Then, the mean-field EOS of a mDPD fluid becomes
\begin{equation}
p_{\rm MF} = \rho k_BT + \alpha_{\rm MF}(A\rho^2 + 2Br_d^4\rho^3).
\label{eq:EOS-mdpd}
\end{equation}
Given proper~$A < 0$ and~$B > 0$, the EOS of Eq.~\eqref{eq:EOS-mdpd} can have a van der Waals loop that enables mDPD to model vapor-liquid coexistence. Ever since its inception, the mDPD model has been successfully applied to investigations of diverse wetting phenomena and multiphase flows. Examples include the two phase flows in a Y-shaped channel~\cite{2011Arienti}, manipulation of liquid droplets using wetting gradient~\cite{2013LiZ} and electrowetting~\cite{2012LiZ}, dynamics of droplets sliding across micropillars~\cite{2015Wang_Chen} and impacting on textured surfaces~\cite{2017Wang_Zhang}, and multiphase flows through nanoporous shales~\cite{2017Xia}, to name but a few.

In the present work, we adopt the mDPD formulation in the form of Eq.~\eqref{eq:FC} to compute the conservative interaction. In a mDPD system consisting of discrete particles, the local density~$\rho_i$ of a mDPD particle~$i$ is computed by the instantaneously weighted summation from its neighbors, i.e., $\rho_i=\sum_{j}w_\rho(r_{ij})$, where the weight function~$w_\rho(r)$ can be any of the smoothing kernels used widely in the smoothed particle hydrodynamics (SPH) method~\cite{2005Monaghan}. Here, we use the three-dimensional Lucy kernel
\begin{equation}
    w_\rho(r) = \frac{105}{16\pi r^3_{c\rho}}\left(1 + \frac{3r}{r_{c\rho}}\right)\left(1 - \frac{r}{r_{c\rho}}\right)^3,
    \label{eq:lucy}
\end{equation}
where~$r_{c\rho}$ is a cutoff radius beyond which~$w_\rho(r)$ vanishes. The factor~${105}/16\pi r^3_{c\rho}$ in Eq.~\eqref{eq:lucy} normalizes~$w_\rho(r)$ so that~$\int_0^\infty d^3\mathbf{r}w_\rho(r) = 1$.

The dissipative and random forces appeared in Eq.~\eqref{equ:gov1} are computed by~\cite{1997Groot}
\begin{align}
  & \mathbf{F}_{ij}^{D} = -\gamma {w_{D}}(r_{ij})(\mathbf{e}_{ij} \cdot \mathbf{v}_{ij})\mathbf{e}_{ij} \ , \label{eq:FD}\\
  & \mathbf{F}_{ij}^{R}\cdot dt = \delta {w_{R}}(r_{ij})\mathrm{d}W_{ij} \mathbf{e}_{ij} \ , \label{eq:FR}
\end{align}
where~$\gamma$ is the dissipative coefficient and~$\delta$ defines the strength of random force,~$\mathbf{v}_{ij} = \mathbf{v}_i - \mathbf{v}_j$ is the relative velocity, and~$w_D(r)$ and~$w_R(r)$ are the weight functions. $\mathrm{d}W_{ij}=\mathrm{d}W_{ji}$ are independent increments of the Wiener process. For a system in thermodynamic equilibrium, the fluctuation-dissipation theorem requires that the dissipative and random forces are related by satisfying~\cite{1995Espanol}
\begin{equation}
\delta^2 = 2 \gamma k_{B}T, \quad w_D(r)=w^2_R(r),
\end{equation}
in which $k_{B}$ is the Boltzmann constant and $T$ the temperature. A typical choice of the weight function is $w_D(r) = w^2_R(r)=(1-r/r_c)^2$ for $r\le r_c$ and zero for $r > r_c$~\cite{2003Warren}.

\subsection{Time Integration}
\begin{figure}[h!]
\centering
\includegraphics[width=0.5\textwidth]{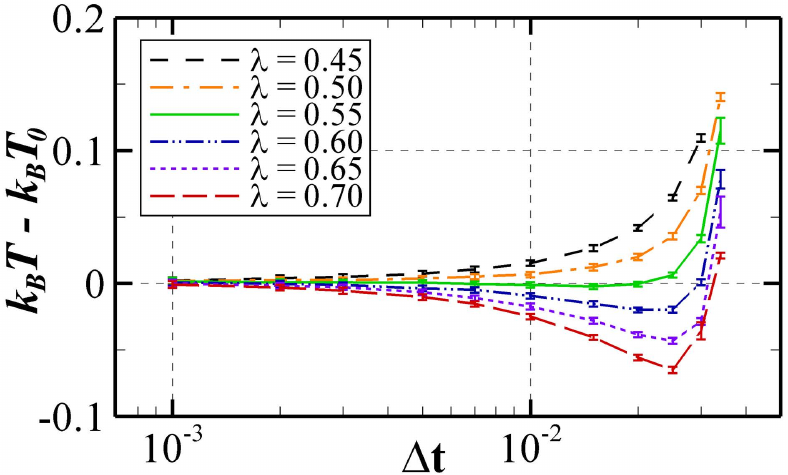}
\caption{Numerical tests to determine the optimum mDPD algorithm. Temperature of the mDPD system as a function of time step~$\Delta t$ (in reduced DPD units) for different values of~$\lambda$, showing the optimal choice for a mDPD system is~$\lambda = 0.55$ (green line). The standard velocity-Verlet algorithm is recovered for~$\lambda = 0.5$.}
\label{fig:mvv}
\end{figure}

In this section, we will clarify a possible misunderstanding of the tunable parameter~$\lambda$ appeared in the modified velocity-Verlet (MVV) algorithm. The velocity-Verlet (VV) algorithm is a standard algorithm for integrating MD and DPD systems due to its symplectic property, numerical stability and simplicity of implementation~\cite{2017Allen}. It first integrates half-step velocity and then one-step position, which are used for the force evaluation, followed by a second half-step velocity integration. The numerical implementation scheme of the VV algorithm is
\begin{align}\label{eq:vv}
\begin{matrix}
 \mathbf{v}(t+\frac{1}{2}\Delta t) &=& \mathbf{v}(t) + \frac{1}{2m}\Delta t\cdot \mathbf{F}(t), \\
 \mathbf{r}(t+\Delta t) &=& \mathbf{r}(t) + \Delta t \cdot \mathbf{v}(t + \frac{1}{2}\Delta t) , \\
 \mathbf{F}(t+\Delta t) &=& \mathbf{F}\left( \mathbf{r}(t+\Delta t), \mathbf{v}(t+\frac{1}{2}\Delta t) \right), \\
 \mathbf{v}(t+\Delta t) &=& \mathbf{v}(t+\frac{1}{2}\Delta t) + \frac{1}{2m}\Delta t \cdot \mathbf{F}(t+\Delta t).
\end{matrix}
\end{align}
However, because the dissipative force in a DPD system depends on velocity, the force calculation in the VV algorithm contains a temporal misalignment between the position and velocity. To this end, Groot and Warren~\cite{1997Groot} introduced a parameter $\lambda$ to make a better prediction of velocity for the force calculation, and proposed the MVV algorithm given by
\begin{align}\label{eq:mvv}
\begin{matrix}
 \mathbf{r}(t+\Delta t) &=& \mathbf{r}(t) + \Delta t \mathbf{v}(t) + \frac{1}{2m}\Delta t^2 \mathbf{F}(t), \\
 \mathbf{\tilde{v}}(t+\Delta t) &=& \mathbf{v}(t) + \lambda \Delta t \mathbf{F}(t), \\
 \mathbf{F}(t+\Delta t) &=& \mathbf{F}\left( \mathbf{r}(t+\Delta t), \mathbf{\tilde{v}}(t+\Delta t) \right), \\
 \mathbf{v}(t+\Delta t) &=& \mathbf{v}(t) + \frac{1}{2m}\Delta t\left[ \mathbf{F}(t) + \mathbf{F}(t+\Delta t) \right].
\end{matrix}
\end{align}
The factor $\lambda$ in MVV was introduced empirically. Groot and Warren~\cite{1997Groot} tested that $\lambda = 0.65$ has the optimal performance for their DPD system. Thereafter, the value $\lambda = 0.65$ was taken as a generally optimal value that has been used in many different DPD systems~\cite{2007Liu,2008Zhao,2009Wu_Guo,2018Zhang}. However, it is important to mention that the optimal value of the empirical parameter~$\lambda$ is {\em system-dependent}. In a mDPD system, whose equation of state is significantly different from a  DPD system, we demonstrate in Fig.~\ref{fig:mvv} that~$\lambda = 0.65$ does not have good performance. A better choice of~$\lambda$ for a typical mDPD simulation is~$\lambda = 0.55$, which allows larger time steps without losing accuracy and numerical stability.

\subsection{Model System and Parameterization}\label{sec:model}
We consider a liquid droplet placed on a solid substrate surrounded by a vapor phase. The system contains three types of interface, i.e., liquid-vapor, solid-vapor and solid-liquid interfaces. Let~$\sigma_{\rm lv}$ represent the liquid-vapor interfacial tension, $\sigma_{\rm sv}$ and~$\sigma_{\rm sl}$ for solid-vapor and solid-liquid interfacial tensions, respectively. The equilibrium contact angle of the droplet is determined from these quantities by the Young equation~\cite{1805Young},
\begin{equation}
\sigma_{\rm lv}\cos \theta_0 = \sigma_{\rm sv} - \sigma_{\rm sl}.
\end{equation}
We refer to the equilibrium contact angle of a droplet on a flat plane surface as the intrinsic contact angle, denoted by~$\theta_0$. In mDPD simulations, the intrinsic contact angle~$\theta_0$ of a droplet on a solid wall is allowed to adjust from hydrophilic to hydrophobic by changing the solid-liquid interfacial tension~$\sigma_{\rm sl}$, which is implemented by varying the mutual attraction between the solid and liquid particles.

\begin{figure}
\centering
\includegraphics[width=0.5\textwidth]{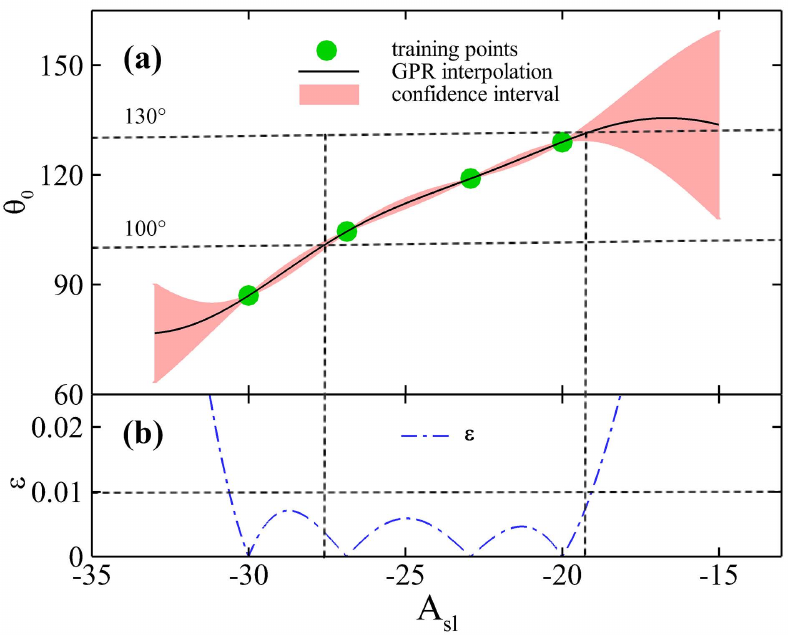}
\caption{Selecting the proper Gaussian Process Regression: (a) GPR-informed mapping from mDPD model parameter~$A_{\rm sl}$ to the intrinsic contact angle~$\theta_0$ based on four training points. The filled circles represent training data, while solid lines are GPR prediction; the shaded area visualizes the prediction uncertainties by the~$95\%$ confidence interval. The contact angles between two dashed lines ($100^\circ<\theta_0<130^\circ$) define the region of interest. (b) The dash-dotted line  shows the magnitude of the acquisition function~$\epsilon(A_{\rm sl})$ for {\it active-learning} compared to an acceptance tolerance~$\delta_{\rm tol}=0.01$.}
\label{fig:ca_Asl}
\end{figure}

In general, constructing a relationship between the intrinsic contact angle~$\theta_0$ and model parameters requires many independent mDPD simulations performed in the parameter space, which is usually a time-consuming process. However, in the present study, only a small range of~$\theta_0$ is of interest because we aim to investigate droplet dynamics only on hydrophobic surfaces. Therefore, we do not have to build a complete parameter-to-$\theta_0$ mapping over a wide range of~$\theta_0$.
To this end, we first define a region of interest, i.e., $\theta_0=100^{\circ}$ to~$130^{\circ}$, and we subsequently apply an {\it active-learning} scheme with a Gaussian Process Regression (GPR) model~\cite{2018Zhao} to minimize the number of necessary mDPD runs and to estimate uncertainties. In particular, a mDPD system of a liquid droplet on a flat solid wall is constructed with parameters~$A_{\rm ss} = A_{\rm ll} = -40.0$, $B = 25.0$, $\gamma = 6.0$, $\delta=18.0$, $r_c = 1.0$, $r_d = r_{c\rho} = 0.75$, $k_BT=1.0$. Because the strength of attractive interaction between solid and liquid particles is determined by the parameter~$A_{\rm sl}$, varying~$A_{\rm sl}$ while keeping the other parameters fixed can tune the solid-liquid interfacial tension~$\sigma_{\rm sl}$, yielding different contact angles. To start the GPR process, two mDPD simulations with~$A_{\rm sl}=-30$ and~$A_{\rm sl}=-20$ are performed to provide two initial training points. By defining an acceptance tolerance~$\delta_{\rm tol}=1\%$, the active-learning scheme can gradually reduce the GPR prediction uncertainties over the region of interest~$\theta_0 = 100^{\circ}$ to $130^{\circ}$ by adding training points adaptively, wherein the location of next sampling point is inferred by an an acquisition function defined as the standard deviation of GPR prediction~$\epsilon(A_{\rm sl})$.
As shown in Fig.~\ref{fig:ca_Asl}, we construct the relationship between the intrinsic contact angle~$\theta_0$ and the parameter~$A_{\rm sl}$, valid for~$100^{\circ}\le\theta_0 \le130^{\circ}$, based on only four mDPD runs, which usually requires more than ten mDPD runs in previous mDPD studies~\cite{2013LiZ,2008Cupelli,2018Lin}. For more technical details of implementing an active-learning scheme with GPR, we refer to a recent work by Zhao et al.~\cite{2018Zhao}.

\begin{table}[t!]
\centering
\begin{tabular}{ccc}
\hline\hline
Quantities & DPD units & Physical units \\\hline
$[m]$  & 1.00 & 2.29$\times$10$^{-15}~kg$ \\
$[L]$ & 1.00 & 2.37$\times$10$^{-6}~m$\\
$[t]$ & 1.00 & 5.58$\times$10$^{-7}~s$\\
$F$ & 1.00 & 9.98$\times$10$^{-11}~N$ \\
$\rho$ & 6.74($\pm$0.04) & 1.16$\times$10$^3~kg/m^3$\\
$\sigma$ & 9.32 ($\pm$0.03) & 6.85$\times$10$^{-2}~N/m$ \\
$\nu$ & 1.16 ($\pm$0.01) & {1.17$\times$10$^{-5}~m^2/s$} \\
\hline\hline
\end{tabular}
\caption{Parameters for mDPD simulations of 60\% glycerol/water solution and units mapping. The symbols $[m]$, $[L]$ and $[t]$ represent mass, length and time units to nondimensionalize the mDPD system. $F$, $\rho$, $\sigma$ and $\nu$ represent force, mass density, surface tension and kinematic viscosity, respectively.}
\label{table:units}
\end{table}

Dimensionless variables are introduced to carry out the mDPD simulations. Three basic physical quantities, i.e., length $[L] = 2.37\times10^{-6}~{m}$, time $[t] = 5.58\times10^{-7}~{s}$ and mass $[m] = 2.29\times10^{-15}~{kg}$, are used to non-dimensionalize the system, leading to a liquid density of $1.16\times10^3~{kg/m^3}$ with a kinematic viscosity of $1.17\times10^{-5}~{m^2/s}$ and a surface tension of $6.85\times10^{-2}~{N/m}$. This corresponds to the physical quantities of the $65\%$ glycerol/water solution at $20~^\circ{\rm C}$~\cite{2012Koichi}. Then, all other physical quantities can be derived from $[L]$, $[t]$ and $[m]$, as shown in Table~\ref{table:units}.

\section{Results and discussion}\label{sec:result}

\begin{figure}[b!]
    \centering \includegraphics[width=0.75\textwidth]{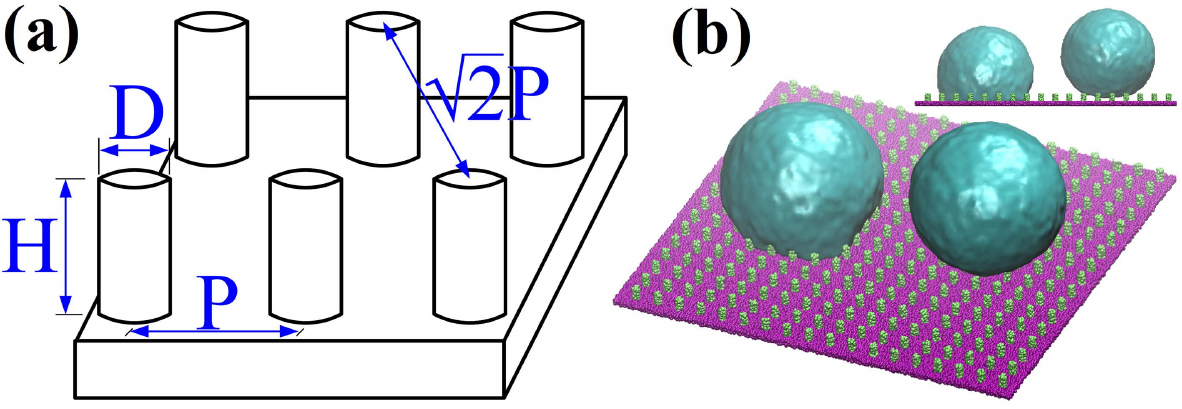}
    \caption{(a) Schematic of the structure of an ideal textured surface, whose microasperities consist of cylindrical pillars with diameter $D = 3.79~\mu m$, height $H = 7.58~\mu m$ and a tunable separation P. (b) Snapshot of mDPD droplets resting on a hydrophobic textured surface in the Wenzel (left) and Cassie-Baxter (right) states, with an inset showing a side view.}
    \label{fig:surface}
\end{figure}

\subsection{The Cassie-Baxter and Wenzel States}
We consider a textured surface with microasperities consisting of cylindrical pillars with diameter $D$, height $H$ and a tunable separation $P$, as shown in Fig.~\ref{fig:surface}(a). The wetting state of an isolated liquid droplet on this rough surface depends on the fraction of solid/liquid interface in their contact area, which has been extensively studied by both experiments~\cite{2009BhushanJY,2009Liu,2011LG} and computational simulations~\cite{2009Koishi,2017ChenZhang}. Wenzel and Cassie-Baxter are the two most widely used models to describe the wetting of droplets on textured surfaces, where a Wenzel state refers to a homogeneous wetting regime shown in Fig.~\ref{fig:FreeEnergy}(a) while a Cassie-Baxter state refers to a heterogeneous wetting regime shown in Fig.~\ref{fig:FreeEnergy}(b). We also present a snapshot of droplets in the Cassie-Baxter and Wenzel wetting states in our mDPD system in Fig.~\ref{fig:surface}(b).
It is generally believed that a Wenzel state can be observed if the maximum gap between pillars ($\sqrt{2}P-D$) becomes large~\cite{2007Nosonovsky,2009BhushanJY}. Let $R$ be the initial radius of a liquid droplet. We define a dimensionless quantity, i.e., the wetting state factor $\alpha$, as a function of ($\sqrt{2}P-D$) and $RH$ in the form of
\begin{equation}\label{eq:alpha}
    \alpha = (\sqrt{2}P-D)^{2}/(RH),
\end{equation}
where a Wenzel state is more likely observed for $\alpha > 1$, while a Cassie-Baxter state is more likely observed for $\alpha < 1$. As depicted in Fig.~\ref{fig:FreeEnergy}(c), there may exist a bidirectional wetting state transition between the Wenzel and Cassie-Baxter states if one breaks the energy barriers $B_1$ and $B_2$.

\begin{figure}[t!]
\centering
\includegraphics[width=0.55\textwidth]{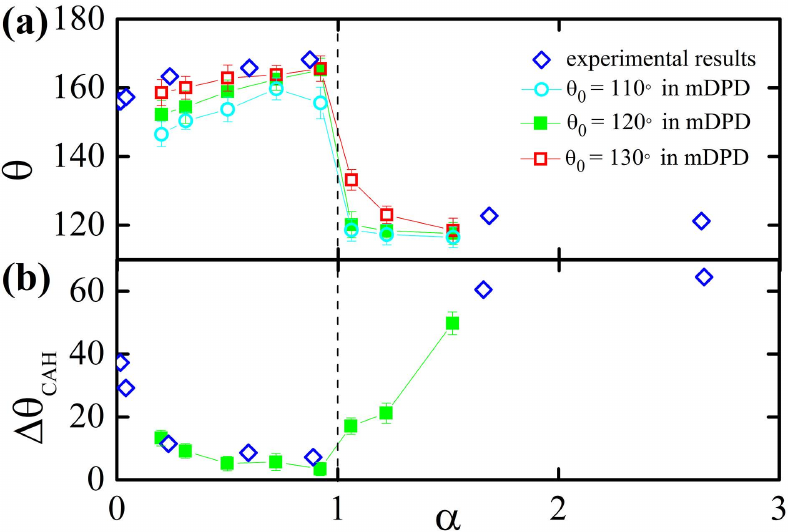}
\caption{Comparison of mDPD results with experiments performed by Bhushan and Jung~\cite{2007BhushanJung} for (a) apparent contact angle~$\theta$ and (b) contact angle hysteresis~$\Delta \theta_{\rm CAH}$ at different wetting state factors~$\alpha$ defined in Eq.~\eqref{eq:alpha}. For apparent contact angles, the mDPD system considers three substrate materials with intrinsic contact angles $\theta_0 = 110^{\circ}$, $120^{\circ}$ and $130^{\circ}$.}
\label{fig:Figure3}
\end{figure}

Given a droplet with radius of $R=44.67~\mu m$, we change the separation $P$ from $8.50~\mu m$ to $18.72~\mu m$ while keeping the pillar diameter $D = 3.79~\mu m$ and the pillar height $H = 7.58~\mu m$ fixed, so that the value of $\alpha$ is varied from $0.20$ to $1.52$. More specifically, we create a solid substrate with cylindrical pillars by cutting selected regions from a thermal equilibrium mDPD system so that the particle distribution in the solid substrate is the same as the particles in liquid phase. With a number density of $\rho=6.74$, the averaged distance between mDPD particles is $\delta r = 1.25~\mu m$. Each cylindrical pillar is made of pre-equilibrated random mDPD particles as in~\cite{2018Li}, and the base of the substrate is a flat plate with a thickness of $3.0~\mu m$. A spherical droplet of $R=44.67~\mu m$ with $189\,040$ mDPD particles is released to the center of the substrate. To imposed the correct no-slip boundary condition on the rough surface, we employ a recently developed boundary method designed for arbitrary-shape geometries~\cite{2018Li}. A mDPD simulation of the droplet is performed for $100$ time units to achieve its thermal equilibrium state, followed by another $100$ time units for statistical analysis. Figure~\ref{fig:Figure3}(a) plots the apparent contact angle $\theta$ of the droplet on the textured surface at different $\alpha$, where we consider three different substrate materials with intrinsic contact angle $\theta_0=110^{\circ}$, $120^{\circ}$ and $130^{\circ}$. For the case of $\theta_0=120^{\circ}$, we find that the droplet on the substrate is in the Cassie-Baxter state with a contact angle $\theta_{\rm CB} > 150^{\circ}$ for~$\alpha < 1$ while in the Wenzel state with a contact angle $\theta_{\rm W}<120^{\circ}$ for~$\alpha > 1$, which is consistent with the experimental results by Bhushan and Jung~\cite{2007BhushanJung}.

When a droplet moves on a solid surface, the triple-phase contact line moves and gives a dynamic wetting response, where the apparent dynamic contact angle differs from Young's contact angle~\cite{1984Joanny}. Taking the contact angle at the front as the advancing contact angle ($\theta_{\rm A}$), the rear as the receding contact angle ($\theta_{\rm R}$), then a contact angle hysteresis $\Delta \theta_{\rm CAH}$ can be defined as the difference of $\theta_{\rm A}$ and $\theta_{\rm R}$, i.e., $\Delta \theta_{\rm CAH} = \theta_{\rm A} - \theta_{\rm R}$~\cite{2006Gao}. The contact angle hysteresis can be experimentally measured by different methods, such as the tilted plate method~\cite{1942Macdougall}, the sessile drop method~\cite{1994Wan} and the Wilhelmy plate method~\cite{2000Abe}. In the present work, we apply a body force on the droplet until it just begins to move over the textured surface to mimic the tilted plate method in experiments. We then compute the advancing and receding contact angles based on the local curvature of the droplet~\cite{2018Zhang,2018ZhaoJY}. Figure~\ref{fig:Figure3}(b) plots the mDPD simulation results of contact angle hysteresis $\Delta \theta_{\rm CAH}$ for different $\alpha$, which are consistent with experimental observations~\cite{2007BhushanJung}. It shows that the droplet in the Cassie-Baxter state has much lower contact angle hysteresis ($\Delta \theta_{\rm CAH}$ is only $3.44^{\circ}$ for $\alpha=0.92$) than the ones in the Wenzel state ($\Delta \theta_{\rm CAH}$ increases to $49.87^{\circ}$ for $\alpha=1.52$). Consequently, the droplets in the Cassie-Baxter state exhibit a much lower adhesion and smaller hydrodynamic resistance and play an important role in self-cleaning of superhydrophobic surfaces.

\begin{figure*}[t!]
\centering
\includegraphics[width=0.8\textwidth]{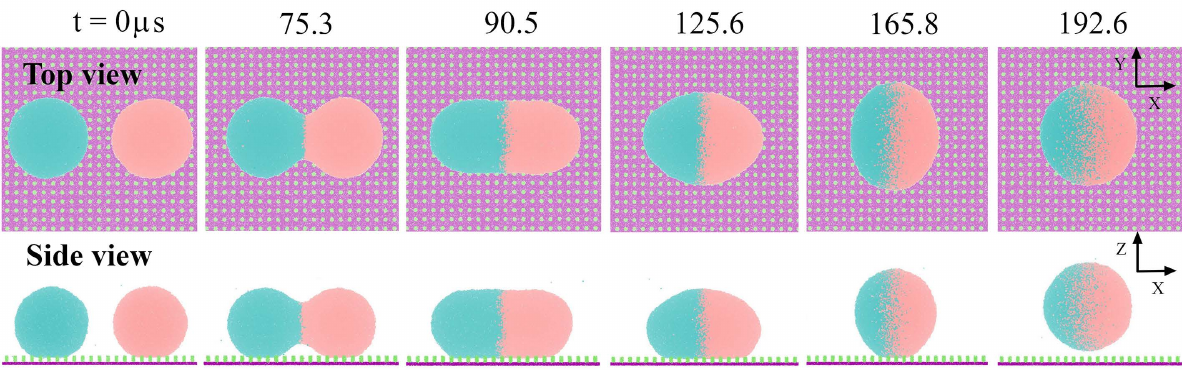}
\caption{Snapshots of coalescence dynamics of two droplets in the Cassie-Baxter state, showing a jumping motion of the coalesced droplet off the textured surface (see also Supporting Information for movie M1). The wetting state factor $\alpha$ for the solid wall is $0.50$ ($P = 11.85~\mu m$), and the apparent contact angle is $\theta_{\rm CB} = 160^{\circ}$.}
\label{fig:Caco_snapshots}
\end{figure*}

\subsection{Coalescence-induced jumping motion}\label{sec:result_1}
We consider the coalescence process of two droplets, both initialized in the Cassie-Baxter state. Two quasistationary liquid droplets of the same size are modeled by $378\,080$ fluid particles, placed on a rough solid surface with an intrinsic contact angle $\theta_0=120^{\circ}$. The textured substrate with size of $237.00~\mu m \times 237.00~\mu m$ and the wetting state factor $\alpha = 0.50$ ($P = 11.85~\mu m$) is built with $85\,328$ frozen particles for the solid base and $17\,346$ particles for $400$ cylindrical pillars.
Periodic boundary conditions are applied in $x$- and $y$- directions, and the no-slip boundary condition for wall surface is implemented using a boundary method designed for arbitrary-shape geometries~\cite{2018Li}.
We first perform a short simulation to relax the two-droplet system to a thermal equilibrium state, having both droplets in the Cassie-Baxter state with an apparent contact angle $\theta_{\rm CB} = 160^{\circ}$, which represents the angle between the apparent solid surface and the tangent
to the liquid-vapor interface~\cite{1999Wolansky}.
Subsequently, we impose a pair of horizontal forces with small magnitude on the droplets so that the two droplets approach each other. The driving force is then removed once the neighboring droplets contact and form a liquid bridge between them, which activates the spontaneous coalescence process powered by the surface energy released upon minimizing the surface area of droplets.
Because the scale of the droplets $R=44.67~\mu m$ is much smaller than the capillary length $l_c=\sqrt{\sigma_{\rm lv}/\rho g}=2.45~mm$, the relevance of gravity for droplet deformation can be safely ignored in our simulations~\cite{2011Wang}.
Figure~\ref{fig:Caco_snapshots} visualizes the time-evolution of the two-droplet system by several snapshots of the coalescence dynamics (see also Supporting Information for the movie M1), where we observe that two neighboring droplets first form a liquid bridge between them, and then coalesce into a bigger one on the textured hydrophobic surface, and finally the resultant droplet jumps off the wall.

\begin{figure}[t!]
\centering
\includegraphics[width=0.5\textwidth]{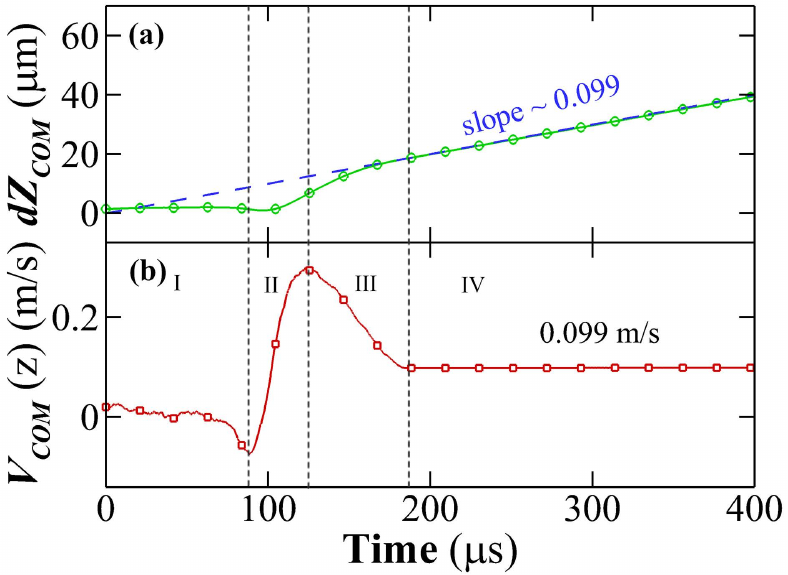}
\caption{Time evolution of (a) the vertical displacement of the center of mass (COM) defined by $dZ_{\rm COM}(t)=Z_{\rm COM}(t)-Z_{\rm COM}(0)$, and (b) the vertical velocity of the center of mass defined by $V_{\rm COM}(t)=d(Z_{\rm COM})/dt$. Contact of droplets starts at 86.5$~\mu s$ and the complete detachment of droplet from the wall surface occurs at 190.5$~\mu s$.}
\label{fig:x_v}
\end{figure}

Taking the two droplets as a group, we study the dynamics of its center-of-mass (COM). Let $Z_{\rm COM}(t)$ be the vertical position of COM at time $t$, $dZ_{\rm COM}(t)=Z_{\rm COM}(t)-Z_{\rm COM}(0)$ be the relative displacement in $z$-direction, and $V_{\rm COM}(t)=d(Z_{\rm COM})/dt$ be the vertical velocity of COM. The time evolutions of $dZ_{\rm COM}(t)$ and $V_{\rm COM}(t)$ are plotted in Fig.~\ref{fig:x_v}, in which we divide the entire coalescence process into four stages for a better interpretation of the simulation results. In stage I ($t < 88.4~\mu s$), a pair of driving forces is imposed on the droplets to make them approach each other, and then a liquid bridge between droplets forms and expands rapidly, until the expanding bridge impacts the solid surface. In stage II ($88.4~\mu s \le t \le 123.9~\mu s$), a high pressure builds up at the bottom of the coalesced droplet because of the non-wetting surface counter effect, leading to an upward acceleration of the coalesced droplet, until the coalesced droplet reaches its maximum velocity. In stage III ($123.9~\mu s < t \le 181.4~\mu s$), the coalesced droplet continues to move upward and starts to detach from the solid surface, while the bottom of the droplet is still in contact with the solid surface, resulting in a deceleration under the effect of adhesive force. Finally, in stage IV ($t > 181.4~\mu s$), the coalesced droplet completely detaches from the solid surface with a jumping velocity $V_J = 9.9~cm/s$.

\begin{figure}[t!]
\centering
\includegraphics[width=0.5\textwidth]{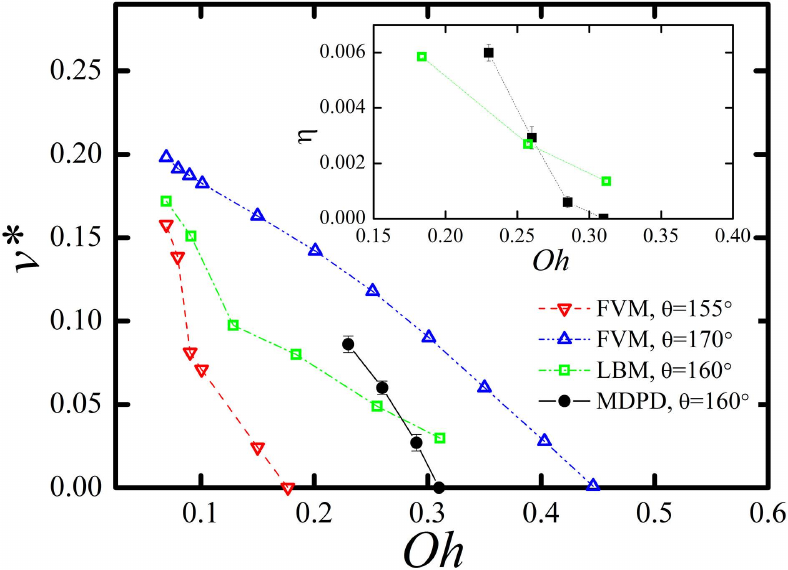}
\caption{Dependence of the dimensionless jumping velocity $v^*$ on the Ohnesorge number $Oh$, with comparison to the results of 3D multiphase LBM simulations by Wang et al.~\cite{2017Wang_Liang} and FVM simulations by Wasserfall et al.~\cite{2017Wasserfall}. The inset shows the energy conversion efficiency for different Ohnesorge numbers.}
\label{fig:v_e_comparison}
\end{figure}

For two equal-sized droplets coalescing on a superhydrophobic surface, several previous experimental and computational works have reported that the jumping velocity of a coalesced droplet $V_J$ increases as the droplet size decreases, which roughly follows an inertial-capillary scaling if viscous dissipation is negligible, i.e., $V_J \propto U=\sqrt{\sigma_{\rm lv}/\rho R}$ where $U$ is defined as the inertial-capillary velocity. However, when the radius of droplet $R$ is smaller than $\sim50~\mu m$, it has been reported that $V_J$ no longer follows the inertial-capillary scaling because of viscous dissipation~\cite{2009Boreyko,2015farokhirad,2017Wang_Liang}. The viscous effects on the droplet coalescence process can be described by the {\it Ohnesorge number} defined by $Oh = \mu /\sqrt{\rho\sigma_{\rm lv}R}$ with $\mu=13.6~cP$ being the dynamic viscosity, which represents the ratio of viscous to surface forces. Here, we define the dimensionless jumping velocity as $v^*=V_J/U$. We perform mDPD simulations with droplet size of $R=44.67~\mu m$, $34.37~\mu m$, $27.68~\mu m$ and $24.17~\mu m$, corresponding to the Ohnesorge number $Oh=0.23$, $0.26$, $0.29$ and $0.31$, respectively. Given $\alpha = 0.50$ unchanged, the corresponding separation $P$ is set to be $11.85~\mu m$, $10.75~\mu m$, $9.92~\mu m$ and $9.45~\mu m$, so that all the four cases are still in the range of the Cassie-Baxter state. Figure~\ref{fig:v_e_comparison} shows the dependence of the dimensionless jumping velocity $v^*$ on the Ohnesorge number, with comparison to the results of 3D multiphase LBM simulations by Wang et al.~\cite{2017Wang_Liang}. The error bars in Fig.~\ref{fig:v_e_comparison} indicate the standard deviations of five independent mDPD simulations. Our results show that the dimensionless jumping velocity $v^*$ decreases with decreasing $R$ (increasing $Oh$), which agrees well with the results of the multiphase LBM simulations~\cite{2017Wang_Liang} and the volume-of-fluid simulations~\cite{2017Wasserfall}, and is consistent with the experimental observations~\cite{2009Boreyko,2013Lv}.

The coalescence dynamics of two identical droplets on a superhydrophobic surface is powered by the released surface energy $\Delta E_s$, which can be computed by~\cite{2017Wang_Liang}
\begin{equation}
\Delta E_s = 2\pi\sigma_{\rm lv}\left[2(1-\cos\theta)+(1- \phi - \phi \cos\theta_0)\sin^2\theta - 2\times\left(\frac{2 - 3\cos\theta + \cos^3\theta}{2}\right)^{2/3}\right]R^2,
\end{equation}
where $\theta$ is the apparent contact angle and $\theta_0$ is the intrinsic contact angle; $\phi = \pi/4 \cdot D^2/P^2$ is the area fraction covered by the cylinders.
We define the energy conversion efficiency $\eta$ as the ratio of the kinetic energy $E_k$ to the released surface energy for the two-droplet coalescence,
\begin{equation}
\eta  = \frac{{{E_k}}}{{\Delta {E_s}}} = \frac{2}{{3{A_0}}}{\left( {{v^ * }} \right)^2},
\end{equation}
where $A_0 = 2(1 - \cos\theta ) + (1 - \phi  - \phi \cos{\theta _0})\sin{^2}\theta - 2(1 - 1.5\cos\theta  + 0.5\cos^3\theta)^{2/3}$. The surface is set to be heterogeneous here with different values of $P$, corresponding to $\phi = 0.08, 0.098, 0.115, 0.126$, $\theta = 160^{\circ}$ and $\theta_0 = 120^{\circ}$. The computed energy conversion efficiency $\eta$ for different $Oh$ is plotted in the inset of Fig.~\ref{fig:v_e_comparison}. We see that the energy conversion efficiency of the coalescence-induced jumping process is inherently inefficient because less than $0.6\%$ of the released surface energy is converted to the translational kinetic energy of the coalesced droplet. As reported by Paulsen et al.~\cite{2012Paulsen}, for $0.1 < Oh < 1$, the coalescence dynamics is in the inertially limited viscous regime. Consequently, as shown in Fig.~\ref{fig:v_e_comparison}, most of the released surface energy in our simulation is dissipated by the viscous effects during the coalescence process, with both the jumping velocity and the energy conversion efficiency decreasing as the Ohnesorge number is increased.

\subsection{Coalescence-induced Wenzel-to-Cassie transition}
We now consider another coalescence process of two droplets, both initialized in the Wenzel state. We use a similar system setup as the previous case in Section~\ref{sec:result_1} with a different pillar separation $P = 8.50~\mu m$. Two equal-sized droplets with radius of $44.67~\mu m$ are placed on a hydrophobic textured surface in the Cassie-Baxter state. As indicated in Fig.~\ref{fig:FreeEnergy}, the Cassie-Baxter state is a metastable state, which can be changed to the stable Wenzel state via a Cassie-to-Wenzel transition by breaking the energy barrier $B_2$. Following Lafuma et al.'s method~\cite{2003Lafuma}, we impose an external body force on each droplet to enforce a Cassie-to-Wenzel transition, and then the external force is removed. We subsequently perform a short mDPD simulation to relax the two-droplet system to their equilibrium Wenzel state. We learned from Fig.~\ref{fig:Figure3} that droplets in the Wenzel state can negate the self-cleaning function of a hydrophobic textured surface because a high adhesive force is produced and a large contact angle hysteresis is generated. Given pillar-textured surface with $P = 8.50~\mu m$, the two droplets with size of $R = 44.67~\mu m$ have a wetting state factor $\alpha = 0.20$, while the coalesced droplet with size of $\hat{R} = 56.28~\mu m$ has a smaller $\alpha = 0.16$. Although both cases have $\alpha < 1$ indicating a favorable Cassie-Baxter state, once the droplet is stuck in the sticky Wenzel state, it cannot change to the Cassie-Baxter state without energy inputs because of the energy barrier $B_1$ shown in Fig.~\ref{fig:FreeEnergy}(c). In this section, we study whether the surface energy released upon two-droplet coalescence is sufficient to create a spontaneous Wenzel-to-Cassie wetting transition.

\begin{figure}[t!]
\centering
\includegraphics[width=0.8\textwidth]{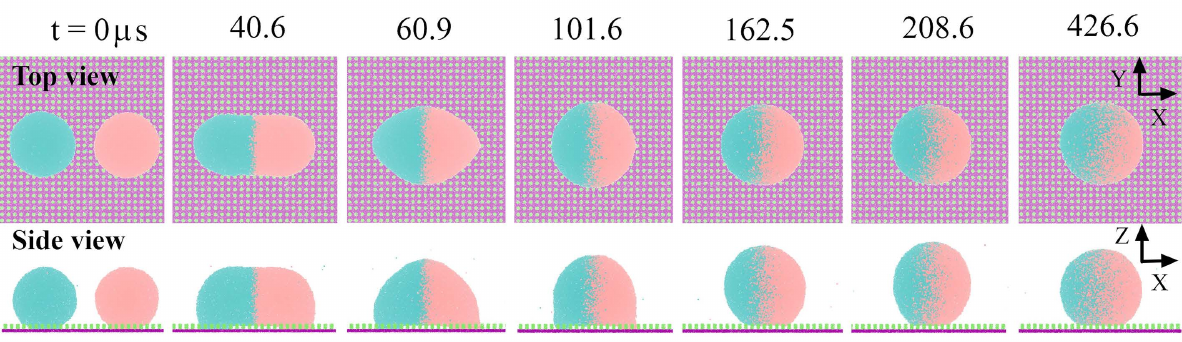}
\caption{Snapshots of coalescence dynamics of two droplets in the Wenzel state, showing a Wenzel-to-Cassie wetting transition powered by the surface energy released upon coalescence of droplets (see also Supporting Information for movie M2).}
\label{fig:weco}
\end{figure}

\begin{figure}[t!]
\centering
\includegraphics[width=0.5\textwidth]{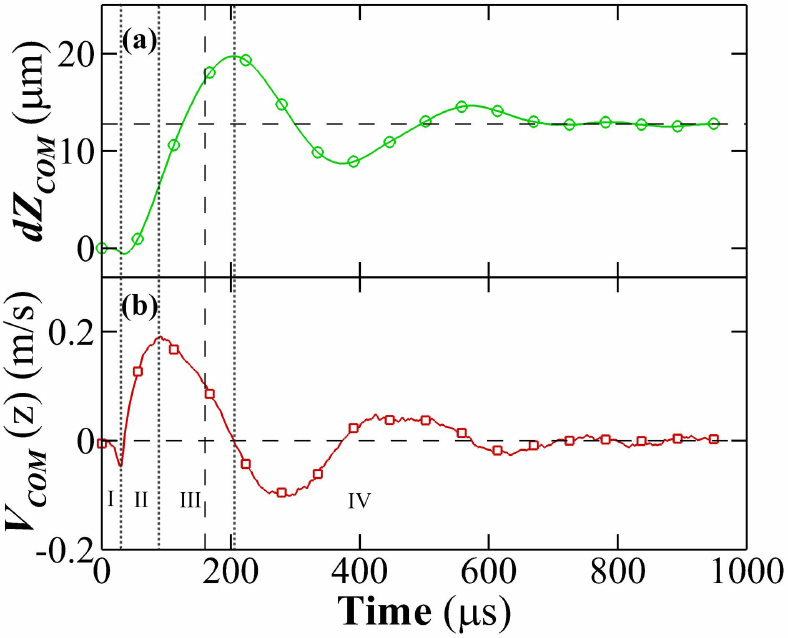}
\caption{Time evolution of (a) the vertical displacement of the center of mass (COM) defined by $dZ_{\rm COM}(t)=Z_{\rm COM}(t)-Z_{\rm COM}(0)$, and (b) the vertical velocity of the center of mass defined by $V_{\rm COM}(t)=d(Z_{\rm COM})/dt$. The vertical dash line shows that the Wenzel-to-Cassie transition is completed after 162.5~$\mu s$.}
\label{fig:x_v_weco}
\end{figure}

Similarly to the previous case, we impose a pair of horizontal forces with small magnitude on the droplets to force them approach each other. The driving force is then removed once the two droplets are in contact and form a liquid bridge between them. Figure~\ref{fig:weco} visualizes the coalescence dynamics of the two droplets, which are initialized in their equilibrium Wenzel state at $t=0$. As the coalesced droplet has a trend of moving upwards, the liquid wetting between the pillars on the surface is pulled out of the gap gradually, and we can observe two obvious pinned anchor points at $t=60.9~\mu s$ in Fig.~\ref{fig:weco}. After $t=162.5~\mu s$, the entire coalesced droplet is out of the gap between pillars, and the Wenzel-to-Cassie wetting transition is completed.

Figure~\ref{fig:x_v_weco} shows the time evolution of the relative displacement of COM $dZ_{\rm COM}(t)$ and the vertical velocity of COM $V_{\rm COM}(t)$, in which we divide the coalescence dynamics into four stages to interpret the results. In stage I ($t < 40.2~\mu s$), the two droplets are in contact and a liquid bridge between droplets forms and expands rapidly, until the expanding bridge impacts the solid surface. In stage II ($40.2~\mu s \le t \le 97.4~\mu s$), a high pressure builds up at the bottom of the coalesced droplet, leading to an upward acceleration of the coalesced droplet, until the coalesced droplet reaches its maximum velocity. The upward movement of droplet pulls the liquid between the pillars out of the gaps. In stage III ($97.4~\mu s < t \le 208.6~\mu s$), the coalesced droplet continues to move upward and pulls the liquid between the pillars out of the gaps, resulting in a deceleration under the effect of adhesive force. Finally, in stage IV ($t > 208.6~\mu s$), the coalesced droplet completes a Wenzel-to-Cassie transition followed by oscillations to dissipate the leftover energy, and finally the droplet is in the low-adhesive Cassie-Baxter state. The spontaneous coalescence-induced Wenzel-to-Cassie wetting transition demonstrates a mechanism that prohibits droplets to remain in the high-adhesive Wenzel state on rough surfaces, which can be interpreted as a possible physical mechanism to maintain the self-cleaning function of hydrophobic rough surfaces.

\begin{figure}[b!]
\centering
\includegraphics[width=0.85\textwidth]{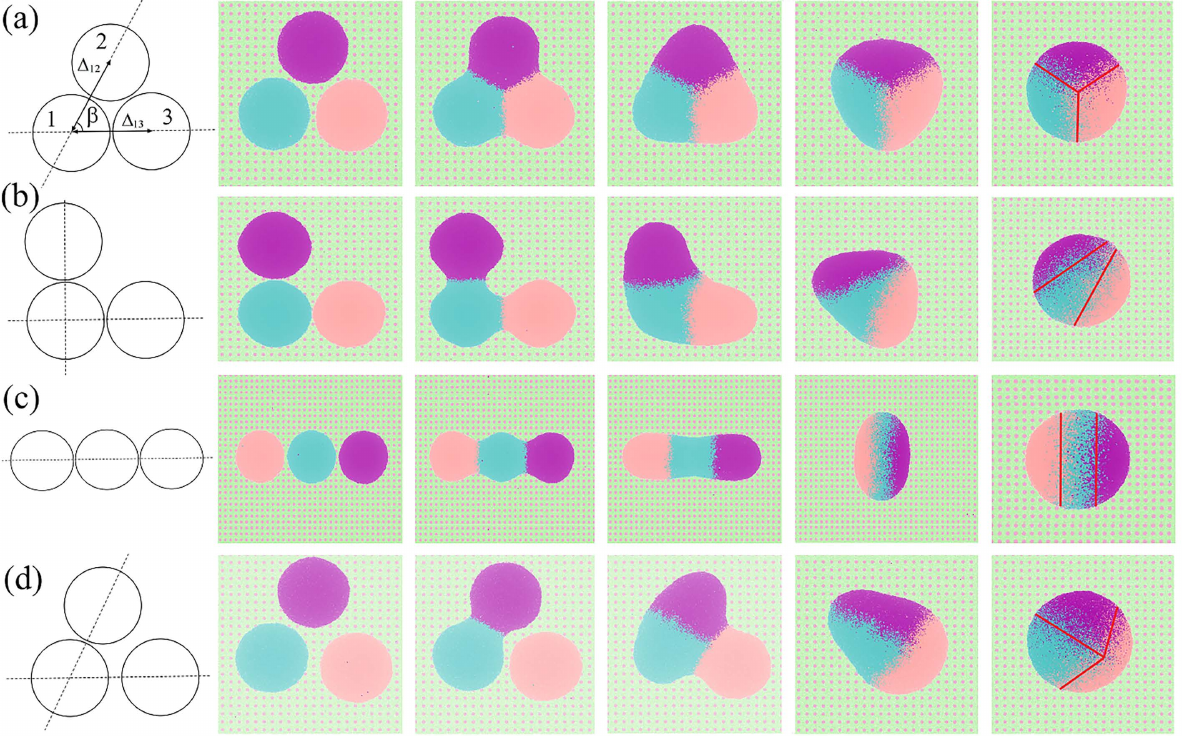}
\caption{Snapshots of initial droplet distribution configurations (top view), in which the top three cases (a)-(c) are in concentrated configurations ($\Delta_{12} = \Delta_{13}$) with $\beta = 60^\circ$, $90^\circ$ and $180^\circ$, while the bottom case (d) is in spaced configuration ($\Delta_{12} \neq \Delta_{13}$). The red lines in the last column show the spatial distribution of liquid components in the coalesced droplet for different configurations (see also Supporting Information for movies M3-M6).}
\label{fig:3dr}
\end{figure}

\subsection{Coalescence of multiple droplets}
We now consider the coalescence process of multiple droplets, which can be also frequently observed in experiments~\cite{2014Watson}. As shown in Fig.~\ref{fig:3dr}, three equal-sized droplets are released onto a textured hydrophobic surface with different overall arrangements. In order to compare with the coalescence-induced jumping of two droplets, we use the same surface morphology as the two-droplet system, i.e., $R = 44.67~\mu m$, $P = 11.85~\mu m$ and $\alpha = 0.50$. Let $O_1$, $O_2$ and $O_3$ be the center of these droplets, $\beta$ be the angle using $O_1$ as the corner, $\Delta_{12}$ be the distance between $O_1$ and $O_2$, and $\Delta_{13}$ be the distance between $O_1$ and $O_3$. We focus on two different kinds of arrangements depending on the relative distance between droplets: concentrated configuration with $\Delta_{12} = \Delta_{13}$ and spaced configuration $\Delta_{12} \neq \Delta_{13}$. Droplets in the concentrated configuration are placed close to each other and coalesce simultaneously because of $\Delta_{12} = \Delta_{13}$, as shown in Fig.~\ref{fig:3dr}(a)-(c). However, droplets in the spaced configuration have different distances, i.e., $\Delta_{12} < \Delta_{13}$, resulting a two steps coalescence process: droplets $1$ and $2$ get into contact first and form a liquid bridge to coalesce, and the droplet $3$ is involved later as it contacts with the expanding liquid bridge of droplets $1$ and $2$, as shown in Fig.~\ref{fig:3dr}(d).

\begin{figure}[t!]
\centering
\includegraphics[width=0.5\textwidth]{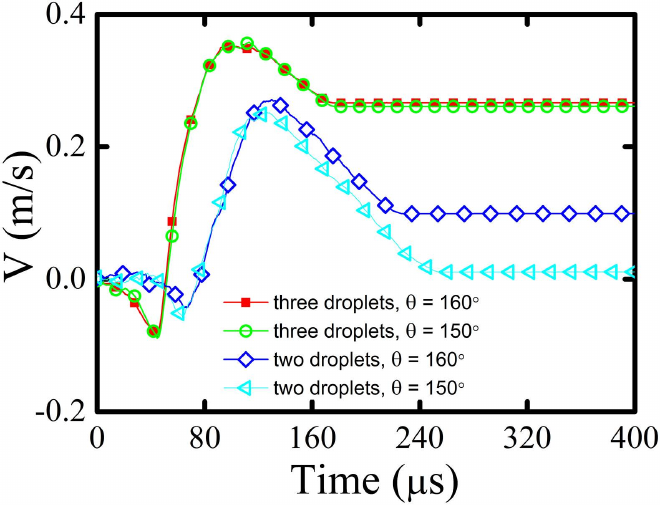}
\caption{\label{fig:2_3_v_comparison}Comparison of center-of-mass velocities during the coalescence process of two-droplet and three-droplet ($\beta = 60^\circ$) systems, where two types of hydrophobic substrates ($\theta_{\rm CB} = 150^\circ$ and $\theta_{\rm CB} = 160^\circ$) with $Oh = 0.23$ ($R = 44.67~\mu m$) are considered.}
\end{figure}

For the case in the concentrated configuration with $\Delta_{12} = \Delta_{13}$, we consider three angles $\beta = 60^{\circ}$ (equilateral triangle), $\beta = 90^{\circ}$ (isoceles right triangle) and $\beta = 180^{\circ}$ (straight line), with top-view snapshots shown in Fig.~\ref{fig:3dr}(a)-(c). The liquids of three droplets are presented with different colors, and hence the snapshots can visualize the coalescence processes for different configurations. We find from the last column in Fig.~\ref{fig:3dr} that the spatial distributions of liquid components inside the coalesced droplet are significantly different, which can be controlled by designing the overall arrangement of droplets and the distance between them. Moreover, a two-step coalescence process presented in Fig.~\ref{fig:3dr}(d) yields an asymmetric distribution of liquid components. These findings may offer new insights for developing new non-contact methods to manipulate liquids inside small droplets based on the initial droplet arrangement.

Coalescence of multiple droplet can release larger surface energy than two-droplet coalescence. For the case in Fig.~\ref{fig:3dr}(a) with $\beta = 60^\circ$, taking the three droplets as a group, we compute its vertical COM velocity $V_{\rm COM}$. Figure~\ref{fig:2_3_v_comparison} plots the time evolution of $V_{\rm COM}$ for two hydrophobic substrates with $\theta_{\rm CB}=150^\circ$ and $160^\circ$, and the comparison to corresponding two-droplet cases. We find that the vertical velocities of both three-droplet coalescence and two-droplet coalescence increase first, followed by a deceleration process, and then end to different jumping velocities. The three-droplet case undergoes a acceleration process last for $61.4~\mu s$, which is similar as the two-droplet case for $58.6~\mu s$. However, the deceleration processes of the three-droplet cases ($71.8~\mu s$) are significantly shorter than the two-droplet cases ($130.2~\mu s$ for $\theta_{\rm CB} = 150^\circ$ and $103.2~\mu s$ for $\theta_{\rm CB} = 160^\circ$). Therefore, the three-droplet coalescence ends up with a much larger jumping velocity than its two-droplet counterpart, i.e., the jumping velocity of three-droplet coalescence for $\theta_{\rm CB} = 150^\circ$ is $V_{\rm J}=0.26~m/s$ compared to $V_{\rm J}=0.01~m/s$ for two-droplet coalescence. It means that adding one more droplet to coalescence may increase the jumping velocity by one order of magnitude. However, we note that the difference between the velocities of three-droplet coalescence for $\theta_{\rm CB} = 150^\circ$ and $\theta_{\rm CB} = 160^\circ$ with $Oh = 0.23$ is not significant, which may because both cases are in the region of $0.1<Oh<1$ where the viscous effect dominates in the coalescence dynamics~\cite{2012Paulsen}.

\section{Conclusions}\label{sec:summary}
We quantitatively investigated the wetting behavior and coalescence dynamics of liquid droplets on mechanically textured hydrophobic surfaces using many-body dissipative particle dynamics (mDPD) simulations. To construct the relationship between the wetting properties of a droplet and the mDPD parameters, we employed an active-learning scheme with a Gaussian process regression model to minimize the number of necessary mDPD runs and to estimate the model uncertainties, which reduces the necessary mDPD runs from more than ten to four. Subsequently, we performed mDPD simulations of liquid droplets (with diameter of $89$ microns) sitting on mechanically textured substrates with microasperities consisting of cylindrical pillars. We defined a dimensionless quantity, the wetting state factor $\alpha$, as a function of the maximum gap between pillars, the droplet radius and the pillar height. By varying the separation distance between pillars, we changed the wetting state factor $\alpha$ from $0.20$ to $1.52$ and quantified the dependence of wetting state on $\alpha$. Simulation results reveal that $\alpha$ can be used to inform the favorable wetting state of a droplet, as the Wenzel state is more likely observed for $\alpha > 1$ while the Cassie-Baxter state is for $\alpha < 1$. We computed the contact angle hysteresis of droplets on textured surface, and verified that droplets in the Cassie-Baxter state have much lower contact angle hysteresis and small hydrodynamic resistance than the ones in the Wenzel state.

We simulated the dynamics of two equal-sized droplets spontaneously coalescing into bigger ones on textured hydrophobic substrates, which is powered by the surface energy released upon droplet coalescence. Given two droplets initialized in a Cassie-Baxter state with an apparent contact angle of~$160^\circ$, we found that the released surface energy is sufficient to cause a jumping motion of droplets off the surface, in which case adding one more droplet to coalescence may increase the jumping velocity by one order of magnitude. We quantified the jumping velocity for droplets with different size, and verified that the jumping velocity of the coalesced droplet decreases as the Ohnesorge number $Oh$ is increased, which agrees well with previous numerical studies and is consistent with experimental observations. We also simulated the coalescence dynamics of two droplets both initialized in the Wenzel state. Instead of a jumping motion, we observed a wetting transition from the high-adhesive Wenzel state to the low-adhesive Cassie-Baxter state. This spontaneous coalescence-induced Wenzel-to-Cassie transition demonstrates a mechanism that prohibits droplets to remain in the highly adhesive Wenzel state on rough surfaces, which can be interpreted as a possible physical mechanism for achieving self-cleaning. Moreover, we studied the coalescence dynamics of multiple droplets, where we changed the overall arrangement of these droplets and observed significantly different spatial distributions of liquid components inside the coalesced droplet. These findings may offer new insights on designing effective biomimetic self-cleaning surfaces by enhancing Wenzel-to-Cassie wetting transition, and additionally, for developing new non-contact methods to manipulate liquid distribution in tiny droplets via multiple-droplet coalescence.

It is worth noting that although the mDPD model was parameterized to capture the surface tension and the viscosity of liquid phase correctly, the surrounding gas phase was represented by the vapor formed by mDPD particles evaporated from liquid phase, which may not be able to produce correct air properties such as pressure and viscous effects. Consequently, the simulations may miss some phenomena relevant to air phase, i.e., a decreasing velocity of droplet due to air friction after it jumps~\cite{2014liuChenS}, and the motion of trapped air underneath the droplet~\cite{2017Attarzadeh}. In these cases, an explicit air phase surrounding the droplet should be included in the mDPD model. Moreover, in the present study, we only considered textured surface, whose microstructure was generated by monosized cylindrical pillars ($3.79~\mu m$ in diameter). Future works should consider biomimetic surfaces with multi-level hierarchical structures, as well as heterogeneous substrates with soft surface coatings for exploring interesting wetting phenomena and coalescence dynamics of droplets.

\section*{Acknowledgements}
This work was supported by the National Natural Science Foundation of China~(Grant No.\ 11872283, 11602133). This work was also supported by DOE PhILMs project~(No.\ DE-SC0019453), and the U.S. Army Research Laboratory and was accomplished under Cooperative Agreement No.\ W911NF-12-2-0023. K. Zhang thanks the China Scholarship Council (CSC) for the financial support (No.\ 201706260015). This research was conducted using computational resources and services at the Center for Computation and Visualization, Brown University.


\end{document}